# Observation of macroscopic valley-polarized monolayer exciton-polaritons at room temperature

N. Lundt[1], S. Stoll[1], P. Nagler[2], A. Nalitov[3], S. Klembt[1], S. Betzold[1], J. Goddard[1], E. Frieling[1], A.V. Kavokin[3,4], C. Schüller[2], T. Korn[2], S. Höfling[1,5] and C. Schneider[1]

[1]*Technische Physik and Wilhelm-Conrad-Röntgen-Research Center for Complex Material Systems, Universität Würzburg, D-97074 Würzburg, Am Hubland, Germany*

[2]*Department of Physics, University of Regensburg, Regensburg D-93040, Germany*

[3]*Physics and Astronomy School, University of Southampton, Highfield, Southampton, SO171BJ, UK*

[4]*SPIN-CNR, Viale del Politecnico 1, I-00133 Rome, Italy*

[5]*School of Physics and Astronomy, University of St. Andrews, St. Andrews KY 16 9SS, United Kingdom*

**In this letter, we address the chiral properties of valley exciton-polaritons in a monolayer of WS$_2$ in the regime of strong light-matter coupling with a Tamm-Plasmon resonance. We observe that the valley polarization, which manifests in the circular polarization of the emitted photoluminescence, is strongly enhanced in comparison to bare WS$_2$ monolayers, and can even be observed under non-resonant excitation at ambient conditions. We study the relaxation and decay dynamics of exciton-polaritons in our device, and present a microscopic model to explain the wave vector-dependent valley depolarization as an interplay of bright and dark states, electron-hole exchange interaction and the linear polarization splitting inherent to the microcavity.**

Excitons which are hosted in two-dimensional atomic crystals of transition metal dichalcogenides (TMDCs) have a variety of intriguing optical properties, which puts them in the focus of advanced light-matter coupling. This includes their enormous exciton binding energies up to 550 meV[1,2], their ultra-fast dipole transitions[3] as well as the materials' unique spin- and valley-related properties[4–7]. The latter are a direct consequence of the broken inversion symmetry of the monolayers in combination with a strong spin-orbit coupling inherited by the transition metal atoms, lifting the polarization degeneracy of the high-symmetry K points at the corners of the Brillouin zone. Thus, the excitons are tagged with a valley index, or valley pseudospin, which can be more robust with respect to depolarization as the exciton pseudospin in conventional III-V semiconductors. Recently, it has been shown that cavity effects, which speed up the relaxation dynamics of excitons, can enhance the time-averaged degree of exciton and trion valley polarization[8,9]. However, in particular in the regime of strong coupling, the hybridization of light and matter leads to a variety of phenomena which have to be addressed to acquire a more profound understanding of these effects. This involves the characteristic quasi-particle dispersion relation which provides various scattering channels, the interplay of bright and dark excitons, as well as the artificial magnetic field provided by the polarization splitting of the cavity resonances. Here, we investigate the relaxation and valley depolarization of valley-tagged exciton-polaritons evolving in a Tamm-plasmon structure[10,11] with an integrated monolayer of $WS_2$ at ambient conditions. Among the TMDCs monolayers, $WS_2$ yields the strongest absorption index, and thus highest oscillator strength[12]. We observe a significant increase in the degree of valley polarization following a non-resonant, circularly polarized excitation as compared to the bare monolayer case. This is attributed to cavity-enhanced transitions between bright and dark states and a speed-up of relaxation in the strong coupling regime, helping to maintain a notable degree of valley polarization of our polariton resonances. This polarization exhibits a distinct dependence on the in-plane wave vector, which we explain as a consequence of k-dependent transfer from dark to bright states and the subsequent population redistribution according to the interplay between the native TE-TM splitting of our microcavity and exchange interactions acting on the polaritons' matter component.

**Monolayer characterization**

The investigated $WS_2$ monolayer was mechanically exfoliated from a bulk crystal and subsequently transferred onto a distributed Bragg reflector (DBR) with a viscoelastic polymer stamp. The DBR is composed of 10 $SiO_2/TiO_2$ layers with thicknesses of 105 nm/65 nm, respectively. The photoluminescence (PL) of the monolayer (on top of the DBR), presented in Fig. 1a, was recorded under 568 nm excitation (continuous wave sapphire laser, 0.7 mW) and ambient condition. The spectrum shows a clear resonance at 618.57 nm/ 2.004 eV with a linewidth (FWHM) of 29 meV, which we attribute the neutral excitonic transition of the Aexcitons. The low-energy shoulder of the peak can be assigned to a trionic transition with an exciton-trion energy splitting of 42 meV, which strongly grows in intensity at lower sample temperatures, in line with previous observations[13]. First, we study the polarization properties of our bare monolayer as a function of the detuning of the pump laser and temperature. Figure 1a depicts a polarization-resolved photoluminescence spectrum of our monolayer exciton recorded at 290 K. The sample was excited with a $\sigma^+$-polarized laser (568nm) and the luminescence is detected in $\sigma^+/\sigma^-$ configurations. The degree of circular polarization (DOCP) is calculated via $P = \frac{I(\sigma^+)-I(\sigma^-)}{I(\sigma^+)+I(\sigma^-)}$, whereas I is the integrated PL intensity. The degree of circular polarization of the emitted light as a function of temperature and excitation energy (532 nm vs. 568nm excitation) is plotted in Fig 1b. While the increase in P with decreasing temperature and an excitation energy closer to the exciton resonance is in excellent agreement with previous findings[14], we emphasize that no significant DOCP can be extracted at ambient conditions.

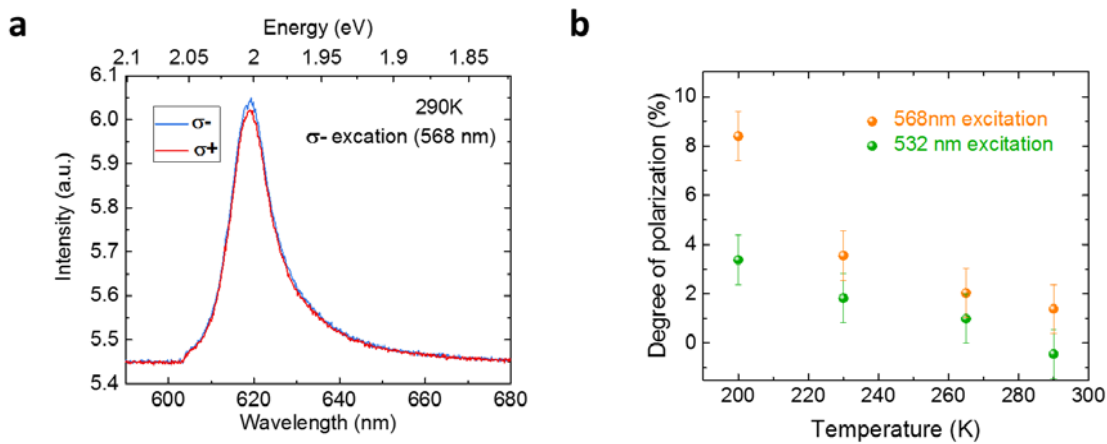

**Figure 1 | Characterization of WS₂ monolayer** a) PL spectra of a monolayer at 290K under $\sigma^+$ polarized excitation at 568nm recorded in $\sigma^+/\sigma^-$ configurations. b) Degree of circular polarization as a function of temperature for 532 nm and 568 nm excitation wavelengths.

**Exciton-Polaritons**

The photonic Tamm structure is completed by capping the $WS_2$ monolayer by 70 nm of PMMA and evaporation 40 nm of silver on top of the PMMA layer (Fig. 2a). A final $SiO_2$ layer was deposited to protect the silver layer from degradation. The bottom DBR supports a very high reflectivity of 99.97 % in a spectral range between 540 nm and 680 nm, and the photonic microstructure features a strong field enhancement close to the metallic interface (see Fig. 2b) at the monolayer location. This strong field enhancement makes such structures particularly interesting to study the physics of exciton-polaritons in solid state systems[15]. The layer thicknesses, illustrated in Fig. 2b as the sequence of the corresponding refractive indices, were designed to promote an optical mode energetically close to the exciton resonance and to spatially overlap with the monolayer. The resonance of the empty cavity was probed in a white light reflectivity measurement presented in Fig. 2c.

The formation of room-temperature exciton-polaritons in our device is confirmed by single-shot angle-resolved photoluminescence measurements in a back-Fourier plane imaging configuration. Utilizing a high magnification (50 x) microscope objective with a numerical aperture of 0.65 allows us to project an in-plane momentum range of up to 5.5 µm$^{-1}$ onto the CCD chip of our spectrometer in this imaging configuration[8]. For polarization measurements we use a λ/4 waveplate to generate $\sigma^+/\sigma^-$ polarized light and analyzed the emitted signal with a rotatable λ/4 waveplate followed by a linear polarizer. We used a beam splitter preserving 98% polarization. In addition, the incident laser was analyzed by a polarimeter at various positions in the optical path in order to ensure a circular degree of polarization more than 99.9%.

The luminescence which we collect from our device is depicted in Fig 2d). It features the typical, distinct dispersion relation of the lower branch of cavity exciton-polaritons, which emerge in the strong coupling regime between excitons and cavity photons. At low k values, this dispersion is dominated by the low mass[16] of two-dimensional cavity photons (approx. $10^{-5}$ $m_e$, $m_e$ being the free electron mass), while its curvature features an inflection point at k~ 3.0 µm$^{-1}$. We can fit the dispersion with a two-coupled oscillator model[15], to extract the Rabi splitting and the exciton-photon detuning of our device. Here, we take advantage of the fact that we can observe the empty cavity dispersion as a faint photoluminescence branch in the background. This PL stems from edge regions of the monolayer and is weakly coupled to the cavity mode. The fit yields a Rabi splitting as large as 80 meV, and an exciton-photon detuning of -55 meV at k=0, which is in well agreement with previous findings on strongly coupled WS$_2$ monolayers in a micro-cavity considering a comparable mode volume[17,18]. The upper polariton branch cannot be observed due to its very low thermal population resulting from the large normal mode coupling strngth.

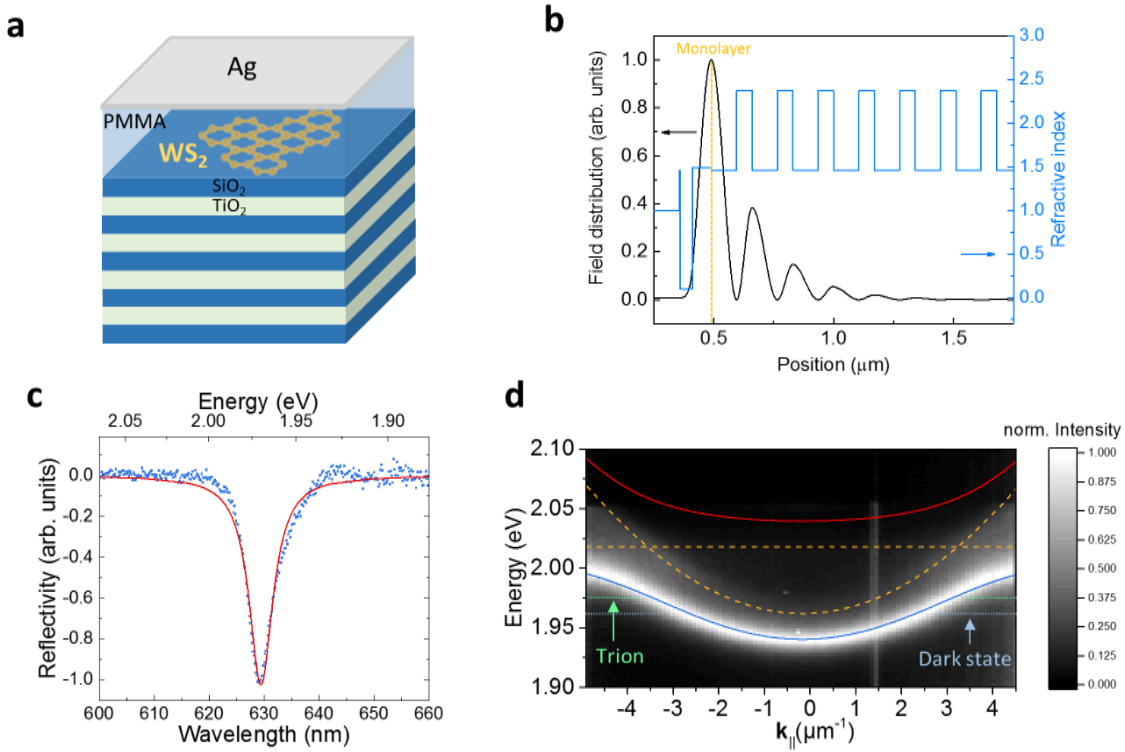

**Figure 2 | Exciton-Polaritons** a) Schematic illustration of the photonic structure with the integrated WS$_2$ monolayer. b) Layer sequence of the photonic structure illustrated by the corresponding refractive indices and the calculated optical field distribution within the photonic structure. The position of the monolayer in the photonic structure is marked as orange, dashed line. c) Measured reflectivity spectrum of the empty cavity at zero in-plane wave vector. d) Exciton-polariton dispersion relation incl. a coupled oscillator fit. The upper and lower polariton branches are drawn in red and blue, respectively, whereas the uncoupled exciton and cavity modes are indicated by orange, dashed lines. Expected energies of trion and dark state are indicated in the background.

We note that due to the large Rabi splitting, our polariton dispersion crosses both the (weakly coupled) trion resonance, as well as the dark exciton (55 meV below the exciton)[19]. Both resonances are indicated as arrows in Fig 2d). While both resonances are in the perturbative regime with our cavity resonance, we believe that the weak coupling conditions can yield a transfer of population to the polariton states.

**Exciton and Polariton Dynamics**

The fitting procedure also provides the light-matter coupling strength, which in turn can be used to assess the radiative lifetime of the WS$_2$ valley excitons. The Rabi splitting yields a direct connection with the exciton oscillator strength and the effective cavity length[20], which can be expressed as:

$$\Omega_R = 2 * \sqrt{\frac{2\Gamma_0 c}{n_c(L_{DBR} + L_C)}}$$

Here, $\Gamma_0$ represents the radiative decay of the excitons, $n_c$ is the effective index of the cavity and $L_{DBR} + L_C$ is the effective cavity length. c is the speed of light in free space. Introducing our system parameters (see supplementary information S1), we yield a radiative decay time being as short as 220 fs. We note that this decay time is in good agreement with previous experimental studies on WSe$_2$ monolayers[3] and theoretical predictions[21]. In our case, we emphasize that our method allows us to solely probe the radiative decay time, as the non-radiative decay channels would not contribute to the light-matter coupling strength. In fact, a calculation of the lower polariton radiative lifetimes in our strongly coupled system yields values that are 3 - 9 times shorter as explained in supplementary S2.

In order to assess the full relaxation dynamics of our system, we perform streak-camera measurements on the bare WS$_2$ monolayer, as well as on the fully built cavity in the strong coupling regime. For this, the sample was excited by a frequency-doubled pulsed fiber laser system (TOPTICA TVIS, pulse length (FWHM) 180 fs, pulse repetition rate 80 MHz) tuned to a central wavelength of 568 nm, coupled into a 100x microscope objective and focused to a spot diameter of less than 1 micron on the sample surface. The PL from the sample was collected using the same objective and coupled into a grating spectrometer, where it was detected using a streak camera coupled to the spectrometer and electronically synchronized with the pulsed laser system. The temporal resolution of this setup (HWHM of the pulsed laser trace) is below 4 ps.

Fig 3a depicts the photoluminescence decay curves of polariton states at high in-plane k-vectors (~ 5 µm$^{-1}$), often referred to as the reservoir, and of the polariton ground state (k=0 µm$^{-1}$). Both time traces exhibit two dominant exponential decay channels and a very weak third decay. The reservoir decay time constants $\tau_1$ and $\tau_2$ of 8.8 ps and 30.3 ps, respectively, are in excellent agreement with the reference measurement on the bare WS$_2$ monolayer (8.2 ps and 31.6 ps, respectively, see supplementary information S3). This is well in line with the highly excitonic character of the polaritons at high k-vectors and previous findings[22,23]. In contrast, characteristic time constants decrease to 6.8 ps and 23.2 ps, respectively, in the polariton ground state. The third decay channel only makes up a minor fraction of the PL intensity (<1%) and is on the order of 100 ps in all measurements. In more detail, Figure 3b and 3c illustrate the decay constants $\tau_1$ and $\tau_2$ as a function of energy which is related to the in-plane wave vector through the dispersion relation. Here, the low-energy end at 1.94 eV represents the polariton ground state, whereas the high-energy tail at 2.01 eV is attributed to emission from the highly excitonic reservoir.

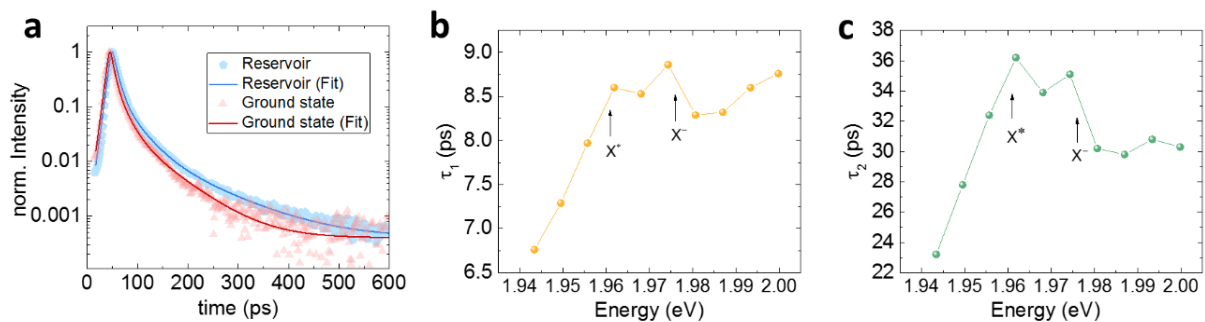

**Figure 3 | Relaxation dynamics** a) Decay curves of the polariton reservoir *(~ 5 µm$^{-1}$)* and the polariton ground state *(~ 0 µm$^{-1}$)* b) Time constant $\tau_1$ as a function of emission energy. c) Time constant $\tau_2$ as a function of emission energy. Energies of the dark state X* and tion X$^-$ are indicated.

Interestingly, both time constants feature a decrease towards lower energies, as well as a slight increase between 1.96 eV and 1.975 eV. In fact, this increase occurs at energies where the polariton dispersion crosses the dark state and the trion resonances, which features a slower decay dynamics[24]. This indicates an indirect pumping mechanism from these weakly coupled states into the polariton states.

Because of the very fast radiative decay the exciton and polariton states inside the light cone, we attribute the decay times as the depopulation dynamics of the original states from which excitons/polaritons scatter into the measured state. These states presumably lie outside the light cone or can be attributed to dark states. Here, we attribute the first channel to fast, phonon-assisted carrier relaxation from outside of the light cone (followed by fast radiative decay) whereas the second, slower decay could be an indication for a transfer from the dark exciton state.

**Polariton Valley Polarization**

In the following, we address the spin- and valley-related properties of our polariton system via polarization-resolved spectroscopy. Here, we inject reservoir excitons non-resonantly via circularly polarized pump lasers with wavelengths of 532 nm or 568 nm, respectively, and measure the DOCP as a function of the polariton wave-vector. The results of this experiment, shown in Fig. 4a (532 nm excitation) and Fig 4b (568 nm excitation), are strikingly different to the case of the bare monolayer, where the DOCP is marginal at room temperature. For the case of 568 nm excitation wavelength, we observe a DOCP as large as 10 % from the polariton ground state at ambient conditions, which is slightly lower for the higher excitation energy (532 nm laser). We furthermore observe a distinct dependency of the DOCP on the in-plane wave vector. The observation of this high circular polarization from the polariton states is, at first sight, surprising, since the system follows the dynamics of the bright exciton reservoir, which we consider as strongly depolarized at 300 K (see Fig 1).

While the relaxation into the bright exciton/exciton-polariton states is almost completely depolarizing, the dark exciton state is considered to maintain the excitation polarization[25,26] as illustrated in figure 4c. While at zero in-plane wave vector the dark exciton has no oscillator strength, it significantly accumulates oscillator strength at large in-plane momenta[19]. Hence, it is reasonable to assume that it can weakly couple to corresponding high-k polariton states (illustrated in Fig. 4c), followed by a radiative decay or a redistribution along the polariton branch. The coupling to the bright state is strongly enhanced by the cavity effect. Thus, we yield a second reservoir at large k$_\parallel$, which is robust against valley depolarization since dark excitons are not subject to strong exchange interactions[25]. See supplementary information S4 for further details.

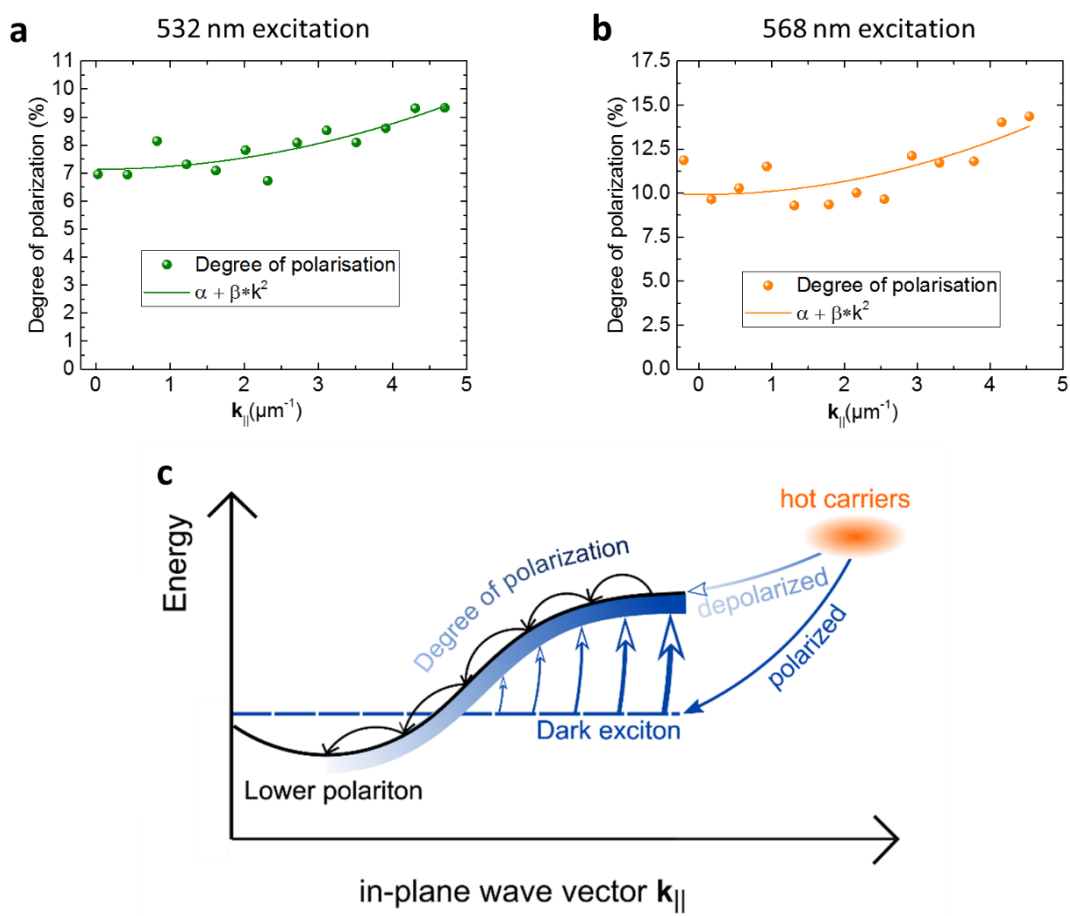

**Figure 3 | Polariton valley-polarization** a) Degree of circular polarization as a function of in-plane wave vector under 532 nm excitation. b) Degree of circular polarization as a function of in-plane wave vector under 568 nm excitation. c) Schematic model for polariton valley polarization.

The transfer from the dark state can be followed by a redistribution of polaritons along the polariton branch. However, in this case the polarized polaritons are again subject to the strong exchange interactions, which can lead to valley depolarization during their redistribution. In the following we provide a description and an analytic expression of the valley depolarization in our polariton system. Here, we consider a gas of free polaritons with parabolic dispersion. We will account for (1) acoustic phonon-assisted energy relaxation, which is dominant below the condensation threshold, and (2) Maialle-Silva-Sham type spin relaxation stemming from the interplay of momentum-dependent effective magnetic field acting on the polariton pseudospin and stochastic elastic momentum scattering. The dynamics of spin relaxation in a polariton gas has been modelled numerically by solving the full set of Boltzmann kinetic equations[18]. In contrast, here we opted for a simplified analytical model with a very limited number of free parameters. We are interested in the analytical expression for the circular polarization of the polariton emission as a function of the in-plane momentum, for the stationary state corresponding to the CW pumping regime.

We discretize the continuous problem and reformulate it in terms of a ladder of discrete quantum states, where direct transitions are only allowed between the neighboring energy levels. We consider phonon assisted inelastic transitions for a polariton with the momentum **p** and neglected nonlinear effects caused by exciton-exciton scattering. In contrast to quantum well based microcavities, where the main contribution to exciton phonon assisted energy relaxation stems from the phonons with momenta oriented along the growth direction[19], the TMDC atomic layers are only coupled to the substrate with the Van der Waals force, so that the excitons only emit in-plane phonons[20]. The characteristic energy lost in a single act of scattering that is taken as an interlevel distance in our model, is then derived from the energy and momentum conservation rules. The rates of polariton transitions between the levels $\alpha_i = \alpha\, p_i^3$ depend on the wave vector via the exciton-phonon matrix element, the exciton Hopfield coefficient, and the density of polariton states.

The spin relaxation rate $\gamma_i$ is governed by the Maialle-Silva-Sham mechanism $\gamma_i = \Omega_i^2 \tau$, where $\Omega_i$ is the value of the effective magnetic field and $\tau$ is the polariton transport time, which we assume constant. The effective field $\Omega_i$, originates from the TE-TM splitting of the photonic polariton component and the long-range part of the electron-hole exchange acting on the excitonic component. In the vicinity of the polariton dispersion minimum the photonic component is dominant, hence, the effective field is given by the TE-TM splitting, which is quadratic in polariton momentum, allowing us to assuyme $\gamma_i = \beta\, p_i^4$.

This yields an analytical expression for the DOCP as a function of the momentum which reads

$$P(p) = P(0) + \frac{\beta}{2\alpha}p^2$$

As shown in fig 4a and 4b, our data can be well fitted by this expression, which confirms our initial assumptions about the important effect of both the electron-hole exchange interaction and the TE-TM splitting in our cavity on the spin-depolarization of valley excitons in TMDC based cavities.

Although our model succeeds to fit the experimental data very well, we note that direct polariton-polariton scattering from reservoir states must be taken into consideration in particular for large pump powers, where the exciton-exciton scattering and the bosonic stimulation become important. This process would result in a speed up in the relaxation dynamics, which is in fact observed for the ground state.

**Conclusion**

In conclusion, our study sheds light into the complex interplay between relaxation, depolarization and decay of valley exciton-polaritons. First, we demonstrate exciton-polaritons in form of strong coupling between photonic microstructure with excitons in WS$_2$ monolayer with a light-matter coupling strength of 80 meV. Based on that we extract a purely *radiative* decay time of excitons which is as short as 220 fs and also calculate the radiative lifetime of the polaritons. We extend these dynamics investigations with time-resolved photoluminescence measurements of the relaxation dynamics. The exciton-polaritons are subject to a speed-up of both the radiative and relaxation decay times, which indicates a non-linear process. This process will be crucial in order to enter the non-linear regime in future experiments. Finally, we investigate the valley polarization of the polaritons, which is significantly increased up to 14% as compared to pure excitons. We attribute this increase to the cavity-enhanced transfer from dark exciton states to the polariton branch, which is followed by a redistribution along the polariton branch. We model this redistribution and quantify the associated depolarization in a way that explains the degree of valley polarization along the polariton branch.

This work has been supported by the State of Bavaria and the ERC (unlimit-2D), as well as the DFG via SFB689, GRK 1570 and KO3612/1-1. We thank S. Tongay for supporting this project. Correspondence and requests for materials should be addressed to Christian Schneider and Nils Lundt (christian.schneider@physik.uni-wuerzburg.de, nils.lundt@physik.uni-wuerzburg.de)

# Supplementary information: Observation of macroscopic valley-polarized monolayer exciton-polaritons at room temperature

N. Lundt[1], S. Stoll[1], P. Nagler[2], A. Nalitov[3], S. Klembt[1], S. Betzold[1], Jacob Goddard[1], E. Frieling[1], A.V. Kavokin[3,4], C. Schüller[2], T. Korn[2], S. Höfling[1,5] and C. Schneider[1]

[1]*Technische Physik and Wilhelm-Conrad-Röntgen-Research Center for Complex Material Systems, Universität Würzburg, D-97074 Würzburg, Am Hubland, Germany*

[2]*Department of Physics, University of Regensburg, Regensburg D-93040, Germany*

[3]*Physics and Astronomy School, University of Southampton, Highfield, Southampton, SO171BJ, UK*

[4]*SPIN-CNR, Viale del Politecnico 1, I-00133 Rome, Italy*

[5]*School of Physics and Astronomy, University of St. Andrews, St. Andrews KY 16 9SS, United Kingdom*

**S1: Polariton system parameters**

In order to access the exciton radiative lifetime, we rewrite

$$\Omega_R = 2 * \sqrt{\frac{2\Gamma_0 c}{n_c(L_{DBR} + L_C)}}$$

to

$$\Gamma_0 = \frac{\Omega_R^2 n_C(L_{DBR} + L_C)}{8c\hbar^2}$$

Now we take $\Omega_R = 82\ meV$, $n_C = 1.4714$ (arithmetic mean of the PMMA layer (70nm, n=1.4915) and the first SiO$_2$ layer (105nm, n=1.458)) and $L_C + L_{DBR} = 235.5\ nm$ (175 nm cavity length + 60.5 nm optical field decay length into the DBR structure, taken from a transfer matrix calculation of the full structure)

## S2: Polariton radiative lifetime

The radiative polariton lifetime $\tau_{polariton}$ can be estimated based the exciton lifetime $\tau_{exciton}$ and cavity lifetime $\tau_{cavity}$. The lifetime is calculated according to (cite deng):

$$\frac{1}{\tau_{polariton}} = (1 - C^2)\frac{1}{\tau_{exciton}} + C^2 \frac{1}{\tau_{cavity}}$$

Where $C^2$ is cavity fraction of the polariton. The exciton lifetime $\tau_{exciton}$ is estimated to be 220 fs (see main text and S1) and the cavity lifetime $\tau_{cavity}$ is calculated according to

$$\tau_{cavity} = \frac{h}{4\pi\,\Delta E}$$

Where h is Plancks constant and $\Delta E$ the cavity linewidth. This yields 20 fs for $\tau_{cavity}$ with a $\Delta E$

of 16.5 meV (equivalent to a Q factor of 120). Figure S1a illustrates the radiative polariton lifetime as a function of the cavity fraction. Figure S1b depicts the lower polariton cavity fraction of our system depending on its in-plane k vector. Taken together the radiative lifetime can be expressed depending in-plane k vector as shown in Figure S1c.

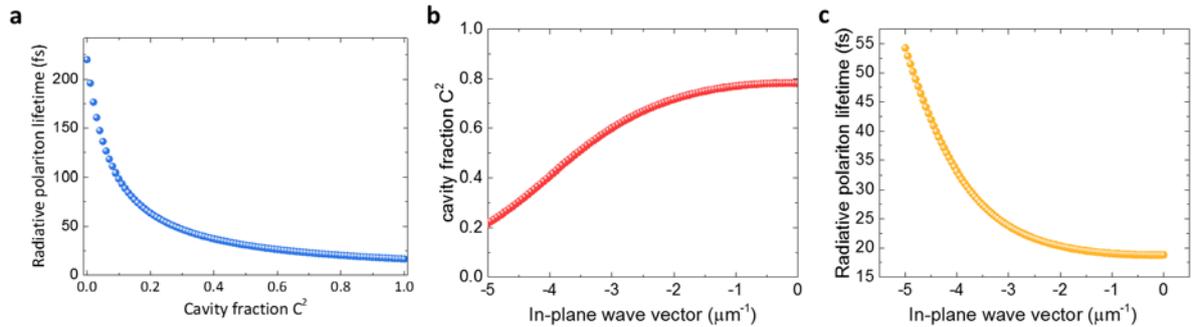

**Figure S2 | Exciton-Polaritons radiavite decay** a)Calculated, radiative decay time of exciton-polaritons as function of the polariton cavity fraction. b) Polariton cavity fraction as a function of its in-plane wave vector. c) Polariton radiative decay time as a function the in-plane wave vector.

**S3: Monolayer dynamics**

We have measured the decay dynamics of a bare WS$_2$ monolayer under ambient conditions in order to compare the dynamics to those in the polariton case. The radiative decay curve of this reference sample is shown in Figure S3. This yields values for $\tau_1$ and $\tau_2$ of 8.2 ps and 31.6 ps, respectively, which are very close to those in the lower polariton case at high k vectors (8.8 ps and 30.3 ps, respectively).

To fit the time traces, we used a convolution of an exponential growth function with the streak camera response function and with three exponential decay functions. The streak camera response function has a line width (FWHM) of 8 ps. As we fit both rise and decay time, the camera temporal resolution is 4 ps (HWHM) for each time constant.

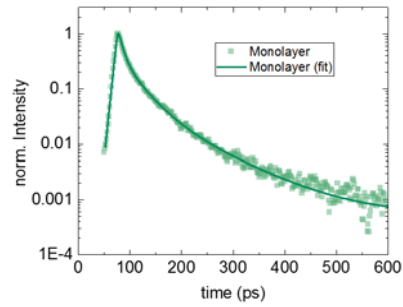

**Figure S3 | WS$_2$ monolayer dynamics** Decay curve of the PL intensity of a WS$_2$ monolayer reference sample

**S4: Transfer from dark to bright state**

In order to estimate the relative coupling efficiency $\gamma$ from the dark state to the polariton state as a function of the in-plane wave vector, we consider the increase of the in-plane dipole moment M($k_{||}$) according to ref.[1] ($M \sim (sin^{-1}(\frac{k_{||}hc}{2\pi E}))^2$, h being Plancks constant, c the speed of light and E the photon energy) and the spectral overlap integral of the spectral distributions of the dark state $S_{dark\ state}(k_{||})$ and of the lower polariton $S_{lower\ polariton}(k_{||})$.

$$\gamma(k_{||}) \sim \int S_{dark\ state}(k_{||}, E) * S_{lower\ polariton}(k_{||}, E)\, dE * M(k_{||})$$

The spectral distribution of the dark state is described by a Gaussian peak with the same linewidth as the bright state (29 meV) and with peak position being 55 meV[1] below the exciton. The spectral distribution of the lower polariton is described by a Gaussian peak with a polariton linewidth depending on its cavity fraction C² (see supplementary S2) and with the peak positon following the dispersion relation (see main text Fig. 2c).

Figure S4 illustrates the result of for $\gamma(k)$, showing that $\gamma$ is significantly increasing with for high in-plane wave vectors.

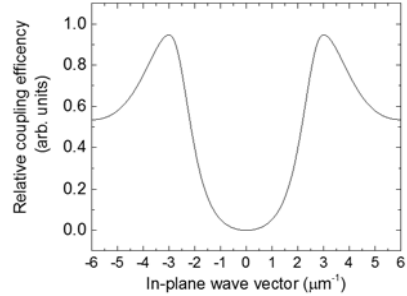

**Figure S4 | Relative coupling efficiency** Relative coupling efficiency of the dark exciton state into the bright polariton state as a function of the in-plane wave vector

## S5: Kinetic model of polariton spin relaxation

We consider a gas of free polaritons with parabolic dispersion, created by a circularly polarized nonresonant pumping at high energies. We will account for (1) acoustic phonon-assisted energy relaxation, which is dominant below the condensation threshold, and (2) Maialle-Silva-Sham type spin relaxation stemming from the interplay of momentum-dependent effective magnetic field acting on polariton pseudospin and stochastic elastic momentum scattering. For the stationary state corresponding to CW pumping we will find an analytical solution for the circular polarization of polariton emission in dependence on the wave-vector.

We discretise the problem by considering possible phonon assisted inelastic transitions of a polariton with momentum $p$. In the limit of slow speed of sound $u \ll p/m$, where $m$ is the polariton effective mass, the maximal energy which can be emitted by the polariton to the thermal bath in one transition reads $\Delta E_{\max} = 2pu$. However, the averaged over scattering angle energy dissipation may be estimated as $\langle \Delta E \rangle = pu$. This argument allows us to set the energy distance of the bosonic ladder $E_{i+1} - E_i = up_{i+1}$, where $E_i$ and $E_i = \sqrt{2mE_i}$ are the energy and momentum value of the $i$-th level respectively. This energy discretization corresponds to homogeneous grid in the momentum value: $p_i = imu$ with the resolution being the momentum of a polariton at the speed of sound.

We introduce the occupation numbers of twice degenerate in spin energy levels $n_i^\pm$ and focus on the linear regime of the bosonic ladder $n_i^\pm \ll 1$. In this case we can neglect the effect of bosonic stimulation and reduce the set of semiclassical Boltzmann kinetic equations to

$$\frac{dn_i^\pm}{dt} = \alpha_{i+1} n_{i+1}^\pm - \alpha_i n_i^\pm - \gamma_i(n_i^\pm - n_i^\mp) - \gamma n_i,$$

where $\gamma$ is the universal rate of polariton decay, $\alpha_i$ and $\gamma_i$ are the level-specific rates of phonon-assisted spin-conserving transitions between the levels and spin relaxation, stemming from elastic momentum scattering within the level. Introducing the level population $n_i = n_i^+ + n_i^-$ and degree of circular polarization $P_i = (n_i^+ - n_i^-)/n_i$, we set the time derivatives in the kinetic equation to zero to find the stability condition:

$$n_{i+1} = n_i \frac{\alpha_i + \gamma}{\alpha_{i+1}},$$

$$P_{i+1} = P_i \left(1 + \frac{2\gamma_i}{\alpha_i + \gamma}\right).$$

From this recursively defined sequence we express

$$P_i = P_0 \prod_{l=0}^{i} \left(1 + \frac{2\gamma_l}{\alpha_l + \gamma}\right),$$

which can be simplified in the case of slow spin relaxation $\gamma_i \ll \alpha_i$:

$$P_i = P_0 \left(1 + \sum_{l=0}^{i} \frac{2\gamma_l}{\alpha_l + \gamma}\right).$$

The interlevel transition rates $\alpha_i$ are given by the product of the squared transition matrix element and the number of states involved in the transition. The exciton-phonon scattering matrix element in Van-der-Waals structures is linear in the phonon momentum on the scale of characteristic polariton wave-vectors. Unlike the QW case, where the main exciton energy relaxation mechanism is due to emission of phonons in the growth direction, here the active layer is mechanically isolated from the substrate, therefore the excitons can only emit in-plane phonons. In turn, the Hopfield coefficient governing the excitonic fraction of polariton also slowly varies in the vicinity of the polariton dispersion bottom, faraway from the anti-crossing point. Finally, the number of states taking part in the transition can be approximated by $\rho(E_i)(E_{i+1} - E_i)$, where $\rho(E_i)$ is the density of states. Since in the two-dimensional case $\rho(E) = const$, the transition rate $\alpha_i = \alpha p_i^3$ is qubic in the polariton momentum.

The spin relaxation rate $\gamma_i$ is governed by the dominating contribution of the Maialle-Silva-Sham mechanism $\gamma_i = \Omega_i^2 \tau$, where $\Omega_i$ is the value of stochastic effective magnetic field and $\tau$ is the polariton transport time, which we assume constant. The effective field value $\Omega_i$ originates from the TE-TM splitting of the photonic polariton component and the long-range part of the electron-hole exchange acting on the excitonic component. In the vicinity of the polariton dispersion bottom the photonic component is dominant, hence, the effective field is given by the TE-TM splitting, which is quadratic in polariton momentum, allowing us to put $\gamma_i = \beta p_i^4$. Assuming fast polariton energy relaxation $\alpha_i \gg \gamma$, we therefore have for the degree of circular polarisation:

$$P_i = P_0 \left(1 + 2\frac{\beta}{\alpha} \sum_{l=0}^{i} p_l\right)$$

Finally, approximating the sum by an integral and coming back to continuous momentum value $p$ we arrive at the expression for the circular polarisation as a function of momentum:

$$P(p) = P(0) + \frac{\beta}{2\alpha} p^2.$$

The neighboring level transition approximation we eployed, which allows us to derive the recurrent relation for the circular polarisation of the emission from the $i$-th level, is based on several assumptions. Firstly, we assume that the characteristic speed of sound $u$ is much slower then the characteristic velocity of polaritons with wave vectors in the region of interest. This implies only one acoustic phonon mode, however, the approximation works in the case of many acoustic modes as long the fastest of them is still slow compared to polariton velocities. Secondly, we base our argument on the existence of the characteristic energy, transferred from the polariton to the phonon bath, depending on the polariton momentum. Finally, we assume the dominance of Maialle-Silva-Sham type mechanism in the spin relaxation. This allows us to describe the simultaneous energy and spin relaxation with two separate terms in the kinetic equations, one of them stemming from the spin-conserving instantaneous energy relaxation steps, and the other one describing elastic process of spin relaxation within each energy level.